# Measurement of Interpersonal Trust in Global Software Development: SLR Protocol

S Zapata[1], JL Barros-Justo[2], G Matturro[3], S Sepúlveda[4]

[1]*Informatics Institute, Universidad Nacional de San Juan, Argentine..*

[2]*School of Informatics (ESEI), Universidade de Vigo, Ourense 32004, Spain.*

[3]*Software Engineering Department, Universidad ORT Uruguay, Montevideo, Uruguay.*

[4]*Computer Science and Informatics Department, Universidad de la Frontera, Temuco, Chile.*

**Content Table**

# 1. Introduction

Advances in Information and Communication Technologies (ICT) have promoted the relationships among people of different geographical zones, creating new technological, cultural and organizational challenges. This has led to the emergence of virtual work teams in the software development business, i.e., groups of developers who work geographically distributed.

Professional software development is complex and normally team based [1]. It is important to recognize the social, collaborative and knowledge sharing aspects which are key to its success [2].

Research has focused on understanding emotions and mood both in software engineering and software development and how the human aspects of a technical discipline can affect final results [3], [4], [5], [6], [7], [8], [9] and improve software quality.

Trust is a fundamental aspect of cooperative works [10] as software development. This is the case in collocated software teams and is even more important when operating in a virtual team environment [11].

Mayer et al. [12] define trust as *"the willingness of a party to be vulnerable to the actions of another party based on the expectation that the other will perform a particular action important to the trustor, irrespective of the ability to monitor or control that other party"*. Trust allows people to participate in risky activities that they cannot control or monitor and even in which they may be disappointed by the actions of others [13], [14].

It is believed that trust is a fundamental factor in determining the success or failure of virtual teams [15], [16], [17]. Given that a high degree of trust yields significant social communication, predictable communication and feedback, we think that trust is a prerequisite for effective mutual adjustment and is therefore necessary for achieving effective coordination of work [18].

Research shows that teams with high degrees of trust are more proactive, more focused on task output, more optimistic, initiate interactions more frequently, and provide more productive feedback [19].

Software development is clearly a human activity and as such is prone to continuous performance improvements. Software measurement is the approach to control and manage the software process by tracking and improving its performance [20].

Software measurement techniques promise to improve the control of the development process, reduce development time and costs, and produce higher quality software [21]. Software measures have been touted as essential resources for improving quality and controlling cost during software development [22], [23]. The level of trust is a valuable metric to make decisions in order to improve the performance of a virtual team.

The purpose of this protocol is to be useful to identify, evaluate and synthesize reported knowledge about the measurement of interpersonal trust (IpT) in virtual software teams. To achieve this goal we applied a research technique known as Systematic Literature Review (SLR). The aim of a SLR is to be as objective, analytical, and repeatable as possible [32]. The SLR conducting phase is shown in Fig. 1.

## 2. Research goal

The need for a systematic review arises from the requirement of researchers to summarize all existing information about some phenomenon in a thorough and unbiased manner [26]. The goal of this SLR is to identify, evaluate and synthesize reported knowledge about the measurement of interpersonal trust (IpT) in virtual software teams (VST). Our interest is focus in interpersonal trust measurement techniques, measured attributes, software development methodologies where these techniques have been applied and usefulness of the measurement.

### 2.1 Research and publication questions

We state the following research questions:

1. RQ1: Which measurement techniques are used to measure IpT in VST?
2. RQ2: Which attributes are used to measure IpT in VST?
3. RQ3: What aspects are affected by IpT measurement-based decisions?.

4. RQ4: Which software development process are reported?

We also considered the following questions related to research methodology and publication details:

1. PQ1: Which research types are reported?
2. PQ2: Which are the main publication venues?

Table 1 shows a detailed description of the research and publication questions.

Table 1. Description of Research and Publication Questions

| Question | Description |
|---|---|
| RQ1 | *Which measurement techniques are used to measure IpT in VST?.*<br><br>Data to be extracted will include: the type of measurement used (objective, subjective, direct, indirect [35]), instruments of measurement used (questionnaire, interview, formula) and the time of the software development process when the measurement was performed (before, during, after). |
| RQ2 | *Which attributes are used to measure IpT in VST?.*<br><br>We want to know which attributes are measured to obtain grades or levels of IpT in VST. Some attributes could be interactions among team members, emotions of team members, personal opinions, degree of knowledge exchange, etc. |
| RQ3 | *What aspects are affected by IpT measurement-based decisions?.*<br><br>IpT measurement-based decisions can apply on process aspects (promote face to face activities), on the team aspects (coach to team members with low interpersonal trust) or on the tools used in VST (promote communication tools that support social interactions). |
| RQ4 | *Which software development process are reported?.* |
| | We aim to identify and classify the software development process where IpT measurement in VST is applied. These process could be agiles (Scrum, Kanban, XP, etc.) or plan-drives (RUP, Prototyping, Waterfall, etc.). |
| PQ1 | *Which research types are reported?*.<br><br>We used the classification proposed by Wieringa et al. [34], including: validation research, evaluation research, solution proposal, philosophical papers, opinions papers and experience papers. |
| PQ2 | *Which are the main publication venues?*.<br><br>We considered scientific Journals, Conferences and Workshops. The extracted data should include: name of the venue and publishing year. |

## 3. Search strategy

According to Wohlin et al. [27], the completeness of the search in a SLR is very important. In the case of a systematic review comparing software engineering technologies (or techniques), completeness is a critical issue [Kitchenham 2015]. Therefore, we will apply complementary searches to achieve maximum completeness of the results of the search process.

A combined search strategy will be used, it include: automated search in databases, manual search in relevant conferences and snowballing search. Fig. 1 shows the application of these techniques during the SLR execution process. The outcome of these search processes will be a set of candidate studies.

### 3.1 Automated search

We will use five scientific online database in this automated search: ScienceDirect, SpringerLink, SCOPUS, IEEEXplorer and ACM Digital Library as electronic data sources. These are the most cited electronic data source in literature review guidelines [28][29][30] [31] [32].

### 3.1.1 Search string

The search string is composed by a set of terms from the RQs. We will follow a four step process to build up the search string:

I. Identify important terms or concepts used in the RQs.
II. Identify terms used in our known set of papers (referent studies, about interpersonal trust in GSD, known by first and second author previously of this work.)
III. Identify synonyms, abbreviations and alternative spellings of terms found in I and II.
IV. Define the search string by joining the synonym terms with the logical operator OR and the set of key terms with AND.

The final search string will be obtained by applying the four step process iteratively with pilot searches, using several combinations of search terms and comparing the results with a set of known papers, a method recommended by Kitchenham and Charters [26].

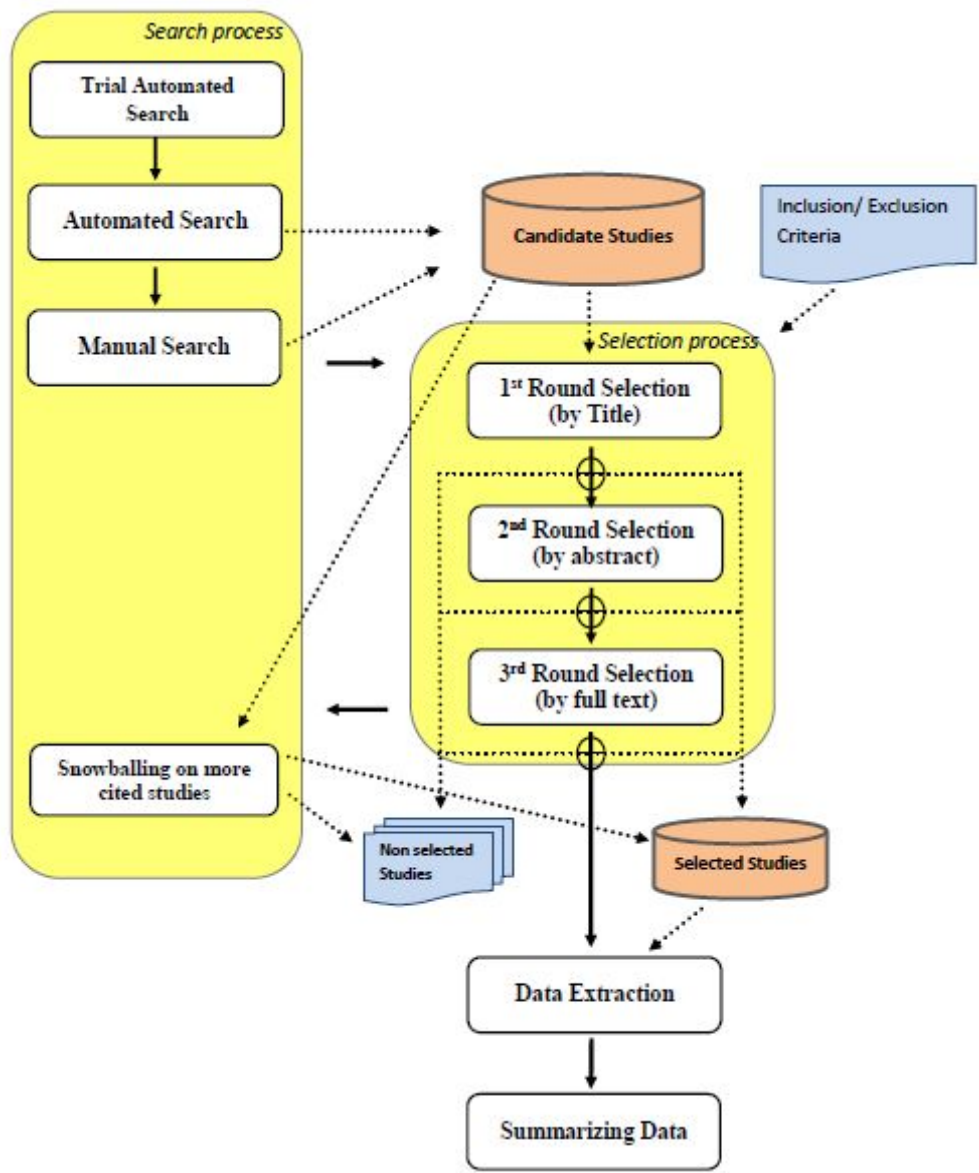

Fig. 1. SLR execution process

## 3.2 Manual search

This complementary search will be performed by manually reviewing the proceedings of the International Conference of Global Software Engineering[1] (ICGSE), International Conference of Software Engineering[2] (ICSE) and Empirical Software Engineering and Metrics Conference[3] (ESEM). Currently, these are the main conferences on Global Software Engineering, general topics of Software Engineering and Software Metrics, respectively.

## 3.3 Snowballing

Although snowballing is a search technique, it will be described in section 5 following the chronological order of processes.

## 4. Selection strategy

The selection strategy consist of a process that is divided into three phases or rounds, as shown Fig. 1. The set of candidate studies, retrieved by the search strategy, will be the input of the selection process, which is grounded on the guidelines proposed in [24].

We will conduct three rounds of the selection process, including selection by title (1st round), selection by abstract (2nd round), and selection by full text (3rd round). If a study is selected in one round, then this paper will not be analyzed in the next rounds. A paper will be selected if it is pertinent for the SLR, i.e. if it meets inclusion criteria.

### 4.1 Inclusion/Exclusion criteria

The purpose of the selection process is to identify a set of relevant papers by applying inclusion/exclusion criteria to the studies retrieved by the searches. The main researcher (first author) defined the inclusion/exclusion criteria, which were reviewed and agreed by the second author. In cases of disagreement the opinion of the third author was imposed. The defined criteria are shown in Table 2.

Table 2. Inclusion/Exclusion Criteria

| # | **Inclusion Criteria** |
|---|---|
| IC1 | Papers written in English. |
| IC2 | Papers published in Journals indexed by the JCR or in International Conferences with a peer-reviewed acceptance system. |

---

[1] icgse.org
[2] https://www.icse2018.org/
[3] http://esem-conferences.org/

| IC3 | Papers focused in *measurement* of *interpersonal trust* in *global software development*. |
|---|---|
| # | **Exclusion Criteria** |
| EC1 | Grey literature (slides presentations, tutorials, forewords, keynote speeches, letters, etc.) |
| EC2 | Short papers (less than 4 pages) |
| EC3 | Duplicate reports of the same study (we consider only the most recent one). |

## 4.2 Distribution of task and deal with disagreements

The selection process will involve three rounds (see Fig. 1): 1st round selection (considering only the paper's title), 2nd round selection (including the abstract) and 3rd round selection (reading the full text). The four authors will conform two teams, of two researchers each, that will work independently by conducting the three selection rounds.

The first team will be integrated by the first author and the third author. The second team will be integrated by the second author and the fourth author. The second team will be integrated by second author and the fourth author. Each researcher will apply the selection process alone. If discrepancies arise over the qualification of each study (include, uncertain or exclude) between researchers of the same team, they will be resolved applying Table 3, following the Wohlin et al. guidelines [27].

Table 3. Criteria to resolve discrepancies

|  |  | **Researcher X (first or third author)** | | |
|---|---|---|---|---|
|  |  | Include | Uncertain | Exclude |
| **Researcher Y (second or fourth author)** | Include | A | B | D |
|  | Uncertain | B | C | E |
|  | Exclude | D | E | F |

A and B: the study is included.
E and F: the study is excluded.
C and D: Both researchers must read in full the paper and qualify it again until obtaining A, B, E or F.

The distribution of candidate studies to analyze in each researchers team is shown in Table 4.

Table 4. Study Distribution

| **Authors** | **Studies to analyse** |
|---|---|

| 1st author | 1st half of candidate studies |
| --- | --- |
| 2nd author | 2nd half of candidate studies |
| 3rd author | 1st half of candidate studies |
| 4th author | 2nd half of candidate studies |

**4.3 Validation of studies selection process**

We will use the Kappa score [33] to validate the studies selection process, see Table 5. If the agreement level among 1st author and 3rd author is "Good/Substantial" or "Very Good/Almost perfect" we will consider that the data extraction process is valid. In other case, we will call a team meeting to discuss the selection process and the reasons for disagreements. The same validation procedure will perform with agreement among 2nd author and 4th author.

Table 5. Interpretation of Kappa score [33]

| **Value of Kappa score** | **Strength of agreement** |
| --- | --- |
| 0 - 0.29 | Poor |
| 0.21 - 0.40 | Fair |
| 0.41 - 0.60 | Moderate |
| 0.61 - 0.80 | Good/Substantial |
| 0.81 - 1.00 | Very Good / Perfect |

In order to minimize threats of completeness in the results of search we will apply snowballing technique, following the Wohlin´s at al. guidelines [27]. The studies that we will obtain with snowballing technique will be added to set of the selected studies.

**5. Snowballing search**

Snowballing is a search approach that uses the reference list of a paper or the citations to the paper to identify additional papers [27]. The snowballing process, shown in Fig. 2, is an iterative process described in [27] and involves two complementary sub-process, backward snowballing and forward snowballing.

In snowballing approach, the first step is to select a start set of studies to use for the snowballing process. This start set must have the pertinent conditions described in [27]. With this start set of papers we can apply backward and forward snowballing.

The backward snowballing consist in using the references of a start set of papers to identify new papers to include to set of selected studies. While forward snowballing refers to identifying new papers based on those studies citing to start set of papers.

Snowballing is a approach quite straightforward to identify relevant papers. It should not necessarily be seen as an alternative to database searches. Different approaches to identifying relevant literature should preferably use to ensure the best possible coverage of the literature [27].

In our case, snowballing will be used as complementary search method and it will be carried after execution of automated and manual search methods. According to Wohlin et al. [27] recommendations we will conform the start set of papers with following characteristics:
- Since our research focus is a specific topic a great number of papers is not necessary. Therefore, to our start set we will choose 30% of the more cited papers from the set of selected studies with automated and manual search.
- Studies will be of different scientific communities.
- Studies will be of different publishers.

## 5.1 Distribution of task and deal with disagreements

The four authors will conform two teams, of two researchers each, that will work independently by conducting the backward and forward snowballing.

The first team will be integrated by first author and third author and will conduce backward snowballing. The second team will be integrate by second author and fourth author and will conduce forward snowballing. If discrepancies arise over the interpretation of the data between researchers of the same team, they will be resolved applying Table 3, following the Wohlin et al. guidelines [27].

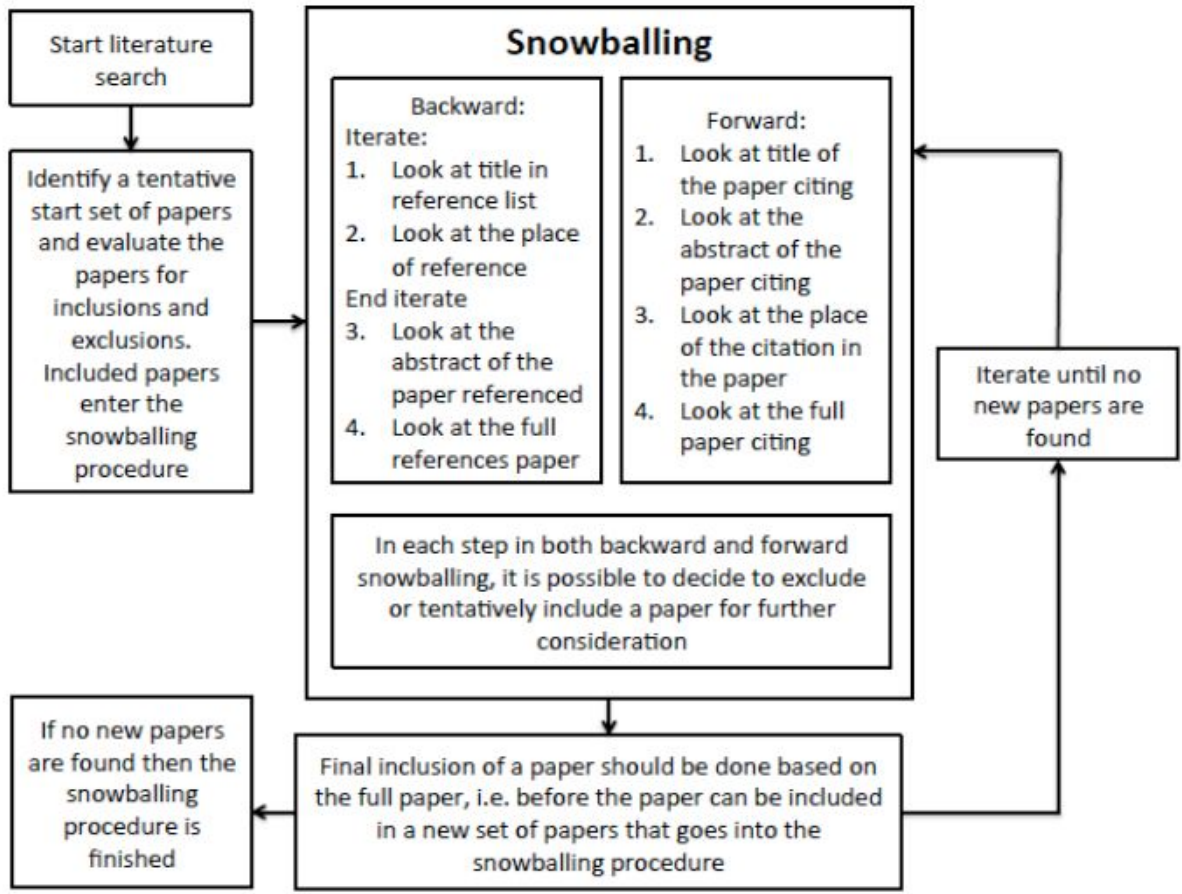

Fig. 2 Snowballing procedure from [27]

## 6. Data Extraction

This part of a protocol defines the data that will be extracted and the process for performing the extraction and validating the data. The data will include publication details for each paper plus the information that is needed to answer the research questions [32].

### 6.1 Data extraction process and deal with disagreements

The extraction of data will be carried out by reviewing the set of selected studies. The full text of each study will be readed to extract data to answer every RQs and PQs listed in Table 1. Data items will be extracted from each paper are shown in Table 6. To keep traceability between processes, an identifier code (Sn) will be assigned to each selected study.

Table 6. Data Items to be extracted from studies

| # | Data Item | Description | Relevant to |
|---|---|---|---|
| **D1** | Title | The title of the paper. | Overview |
| **D2** | Authors List | The full name of all authors of the paper. | Overview |
| **D3** | Abstract | The abstract of the paper. | Overview |
| **D4** | Year | Year when the paper was published. | Overview |
| **D5** | Type of research | The six types of research proposed by Wieringa [34]: Validation research, Evaluation research, Solution Proposal, Philosophical papers, Opinions papers, Experience papers. | PQ1 |
| **D6** | Venue | The name of the venue where the study was published. | PQ2 |
| **D7** | Type of Venue | Journal or Conference | PQ2 |
| **D8** | Type of measurement | Proposed by Wohlin [35]: Direct/Indirect, Objective/Subjective, No report. | RQ1 |
| **D9** | Measurement Instrument | Interview/Questionnaire/Counter/Another/ No report. | RQ1 |
| **D10** | Measurement Time | Before/During/After respect to software process. | RQ1 |
| **D11** | Measured attributes | Interactions/Emotions/knowledge exchange/ Member biography/ Member Opinion/ Another/ No report. | RQ2 |
| **D12** | Affected aspects by IT metric based decisions. | Software process/ team / tools / Another / No report. | RQ3 |
| **D13** | Type of software development process | Proposed by Boehm [36]: Plan-drive/ Adaptive/ Agile methodologies/ Another/ Several/ No report. | RQ4 |
| **D14** | Software development process name | Scrum/XP/Kanban/Waterfall/RUP/Another/No report. | RQ4 |

For systematic reviews, data extraction may be iterative since important trends and ways of categorising papers may only become evident as individual papers are read [32]. If new research questions arise during the data extraction, we will redo the process in order to including these research questions.

To avoid individual bias we define a process to extract data. This process consist in:

1. The selected studies set will be reviewed, to extract data, by 2nd author and 3th author in independent way. Therefore, each study will be reviewed by two researchers.
2. The first author will integrate results of step 1. In case of discrepancy, the 4th author criterion will be imposed based on the 2nd and 3th authors technical opinions.

**6.2 Data extraction form**

The data will be recorded using a spreadsheet with the format shown in Table 7.

Table 7. Data extraction form

| Studies | D1 | D2 | ……………….. | D13 | D14 |
|---|---|---|---|---|---|
| S1 | | | | | |
| S2 | | | | | |
| ………. | | | | | |
| ………. | | | | | |
| Sn | | | | | |

**6.3 Validation of data extraction**

Similar to the quality evaluation process, the agreement between independent assessors is used to assess the validity of the data extraction process. If the agreement for individual primary studies is very low, the data form may not be well understood, so there should be a contingency plan ready [32].

We will use the Kappa score [33] to validate the data extraction process. If the agreement level among 1st author and 2nd author is "Good/Substantial" or "Very Good/Almost perfect" we will consider that the data extraction process is valid. In other case, we will call a team meeting to discuss the data form and the reasons for disagreements.

**7. Summarizing data**

In this section we will present the extracted data from the set of selected studies, for each data element defined in Table 6. The goal should be to classify the findings in interesting ways and to present summaries using a variety of tabular or graphical forms [32]. Six approaches to visualize the classified data have been identified: heatmap, Venn Diagram, Bubble plot, Bar plot, Pie diagram, Line Diagram. The most common approaches are bar plots and bubble plots [24].

According to the resulting data of extraction process, we will use tabular or graphical forms. Only as preliminary proposal the Table 8 shows how each data item of Table 6 would be visualized.

Table 8 Forms for data visualization

| # | Data Item | Form |
|---|---|---|
| **D1** | Title | Tabular form. |
| **D2** | Authors List | Tabular form (or graphical form depending amount of identified authors) |
| **D3** | Abstract | It not will be visualized. |
| **D4** | Year | Bar or bubble plot. |
| **D5** | Type of research | Bar plot. |
| **D6** | Venue | Tabular form (or graphical form depending amount of identified venues) |
| **D7** | Type of Venue | Bar plot. |
| **D8** | Type of measurement | Bar plot. |
| **D9** | Measurement Instrument | Bar plot. |
| **D10** | Measurement Time | Bubble plot. |
| **D11** | Measured attributes | Bar plot. |
| **D12** | Affected aspects with IT metric based decisions. | Bar plot. |
| **D13** | Type of software development process | Bar Plot. |

| **D14** | Software development process name | Tabular form (or graphical form depending amount of identified process) |

## 8. Threats to validity

Petersen and Gencel [37] reviewed existing validity classification schemes and discussed their applicability to software engineering. The following types of validity should be taken into account: descriptive validity, theoretical validity, generalizability and interpretive validity [24]. We will aim mitigate these threat types with several actions.

### 8.1 Descriptive validity

Descriptive validity is the extent to which observations are described accurately and objectively [24]. To objectify the data extraction process we will use a form (see Table 7) and we will apply a process of validation, described in section 6.3, which involve redundant revision.

### 8.2 Theoretical validity

Theoretical validity is determined by our ability of being able to capture what we intend to capture [24]. Four different activities can be affected, the identification and selection of papers (missed studies) and data extraction and classification (researcher's bias).

The actual protocol proposes to use six scientific online database for the automatic search of the primary papers. Additionally, manual search and a snowballing search will be applied to complement the results. This search strategy will reduce the risk of missing some important work.

To reduce researchers bias during the identification and selection process a procedure of redundant revision will be used, see section 4. A Kappa score based validation process will be applied, see section 4.3. Similarly, to minimize bias in data extraction and identification process a redundant revision will be applied, see section 6.3.

In order to control the selected studies quality only papers published in journals indexed by the JCR or in international conferences with a peer-review acceptance system will be included. Gray literature will be excluded, see Table 2.

### 8.3 Generalizability

Generalizability is the extent to which the effects observed in a study are applicable outside of the study [32]. The actual protocol involves a research goal which has a well defined study context, Global Software Development. The selected studies will report data respect to GSD. Hence, the SLR conclusions will be generalizable to this context.

**8.4 Interpretive validity**

Interpretive validity is achieved when the conclusions drawn are reasonable given the data, and hence maps to conclusion validity. A threat in interpreting the data is researcher bias [24]. To reduce threats two experienced researchers will elaborate the conclusions and a statistical researcher will review the results.

**9. Conclusions**

The actual work proposes a SLR based protocol to identify, evaluate and synthesize evidence about Measurement of Interpersonal Trust in a context of Global Software Development. The protocol provides details respect to research goals, searching and selection process of studies, data extraction data and data visualization. Also, it includes activities to minimize the bias threats during process conducting.

We aim that this document provides all important details to researchers about our SLR process.